\documentclass[%
a4paper,
prd,
twocolumn,
superscriptaddress,
preprintnumbers,
nofootinbib,
nobibnotes,
amsmath,amssymb,
aps,
fleqn,
floatfix
]{revtex4-2}
\usepackage{amsfonts,graphics,color}
\usepackage[english]{babel}
\usepackage{graphicx}
\usepackage{subfigure}
\usepackage{amsmath}
\usepackage{dcolumn}
\usepackage{bm}
\usepackage{comment}
\usepackage{multirow}
\usepackage{xcolor}
\definecolor{xlinkcolor}{cmyk}{1,1,0,0}
\usepackage[bookmarks=true, pdfnewwindow=true, colorlinks=true, linkcolor=xlinkcolor, citecolor=xlinkcolor, filecolor=xlinkcolor, urlcolor=xlinkcolor, final=true]{hyperref}

\usepackage[normalem]{ulem}
\definecolor{myblue}{rgb}{0.05,0.1,0.5}

\begin{document}

\preprint{INR-TH-2025-006}

\title[CASH haloscope proposal]{Search for dark-matter axions beyond the quantum limit:\\ the Cosmological Axion Sarov Haloscope (CASH) proposal}

\author{Andrey L. Pankratov}
\thanks{Corresponding author, e-mail: alp@ipmras.ru}
\affiliation{Nizhny Novgorod State Technical University n.~a. R.~E.~Alekseev, Nizhny Novgorod, Russia}
\affiliation{Institute for Physics of Microstructures of RAS, Nizhny Novgorod, Russia}
\author{Pavel A. Belov}
\affiliation{School of Physics and Engineering, ITMO University, Saint  Petersburg 197101, Russia}
\author{Eduard E. Boos}
\affiliation{D.V. Skobeltsyn Institute of Nuclear Physics, Lomonosov Moscow State University, Moscow 119991, Russia}
\author{Alexander S. Chepurnov}
\affiliation{D.V. Skobeltsyn Institute of Nuclear Physics, Lomonosov Moscow State University, Moscow 119991, Russia}
\author{Alexander V. Chiginev}
\affiliation{Nizhny Novgorod State Technical University n.~a. R.~E.~Alekseev, Nizhny Novgorod, Russia}
\affiliation{Institute for Physics of Microstructures of RAS, Nizhny Novgorod, Russia}
\author{Alexander V. Derbin}
\affiliation{Petersburg Nuclear Physics Institute NRC KI, Gatchina 188300, Russia}
\affiliation{National Research Centre Kurchatov Institute, Moscow 123182, Russia}
\author{Ilia S. Drachnev}
\affiliation{Petersburg Nuclear Physics Institute NRC KI, Gatchina 188300, Russia}
\author{Lev V. Dudko}
\affiliation{D.V. Skobeltsyn Institute of Nuclear Physics, Lomonosov Moscow State University, Moscow 119991, Russia}
\author{Dmitry S. Gorbunov}
\affiliation{Institute for Nuclear Research of the Russian Academy of Sciences, 60th October Anniversary Prospect 7a, Moscow 117312, Russia}
\affiliation{Moscow Institute of Physics and Technology, 141700 Dolgoprudny, Russia}
\author{Maxim A. Gorlach}
\affiliation{School of Physics and Engineering, ITMO University, Saint  Petersburg 197101, Russia}
\author{Vadim V. Ivanov}
\affiliation{Nizhny Novgorod State Technical University n.~a. R.~E.~Alekseev, Nizhny Novgorod, Russia}
\affiliation{Institute for Physics of Microstructures of RAS, Nizhny Novgorod, Russia}
\author{Leonid V. Kravchuk}
\affiliation{Institute for Nuclear Research of the Russian Academy of Sciences, 60th October Anniversary Prospect 7a, Moscow 117312, Russia}
\author{Maxim V. Libanov}
\affiliation{Institute for Nuclear Research of the Russian Academy of Sciences, 60th October Anniversary Prospect 7a, Moscow 117312, Russia}
\affiliation{Moscow Institute of Physics and Technology, 141700 Dolgoprudny, Russia}
\author{Michael M. Merkin}
\affiliation{D.V. Skobeltsyn Institute of Nuclear Physics, Lomonosov Moscow State University, Moscow 119991, Russia}
\author{Valentina N. Muratova}
\affiliation{Petersburg Nuclear Physics Institute NRC KI, Gatchina 188300, Russia}
\author{Alexander E. Pukhov}
\affiliation{D.V. Skobeltsyn Institute of Nuclear Physics, Lomonosov Moscow State University, Moscow 119991, Russia}
\author{Dmitry V. Salnikov}
\affiliation{Institute for Nuclear Research of the Russian Academy of Sciences, 60th October Anniversary Prospect 7a, Moscow 117312, Russia}
\affiliation{Lomonosov Moscow State University, 1-2 Leninskie Gory, Moscow 119991, Russia}
\author{Petr S. Satunin}
\affiliation{Institute for Nuclear Research of the Russian Academy of Sciences, 60th October Anniversary Prospect 7a, Moscow 117312, Russia}
\author{Dmitrii A. Semenov}
\affiliation{Petersburg Nuclear Physics Institute NRC KI, Gatchina 188300, Russia}
\author{Alexander M. Sergeev}
\affiliation{Gaponov-Grekhov Institute of Applied Physics of the Russian Academy of Sciences, Nizhny Novgorod 603950, Russia}
\affiliation{National Center for Physics and Mathematics, Bldg.~3, Parkovaya 1, Sarov  607182, Russia}
\author{Maksim I. Starostin}
\affiliation{Lomonosov Moscow State University, 1-2 Leninskie Gory, Moscow 119991, Russia}
\affiliation{Branch of Lomonosov Moscow State University in Sarov, 607328, Russia}
\author{Igor I. Tkachev}
\affiliation{Institute for Nuclear Research of the Russian Academy of Sciences, 60th October Anniversary Prospect 7a, Moscow 117312, Russia}
\author{Sergey V. Troitsky}
\affiliation{Institute for Nuclear Research of the Russian Academy of Sciences, 60th October Anniversary Prospect 7a, Moscow 117312, Russia}
\affiliation{Lomonosov Moscow State University, 1-2 Leninskie Gory, Moscow 119991, Russia}
\author{Maxim V. Trushin}
\affiliation{Petersburg Nuclear Physics Institute NRC KI, Gatchina 188300, Russia}
\author{Evgenii V. Unzhakov}
\affiliation{Petersburg Nuclear Physics Institute NRC KI, Gatchina 188300, Russia}
\author{Maxim M. Vyalkov}
\affiliation{Lomonosov Moscow State University, 1-2 Leninskie Gory, Moscow 119991, Russia}
\affiliation{Branch of Lomonosov Moscow State University in Sarov, 607328, Russia}
\affiliation{National Center for Physics and Mathematics, Bldg.~3, Parkovaya 1, Sarov  607182, Russia}
\affiliation{Russian Federal Nuclear Center -- All-Russian Scientific Research Institute of Experimental Physics, Sarov, Russia}
\author{Arkady A. Yukhimchuk}
\affiliation{Russian Federal Nuclear Center -- All-Russian Scientific Research Institute of Experimental Physics, Sarov, Russia}

\date{Submitted to \textit{Physical Review D} on June 18, 2025}

\begin{abstract}
\end{abstract}
\maketitle

\section{Introduction}
\label{sec:intro}
Axions and axion-like particles (ALPs) are hypothetical pseudoscalar particles predicted in several extensions of the Standard Model of particle physics (SM), including models with flavor symmetries, supersymmetry, extra dimensions, string theory, etc., see e.g.\,\cite{Berezinsky:1993fm,Gorbunov:2000th,deAlwis:2005tf,Pilaftsis:2008qt,Arvanitaki:2009fg,Gorbunov:2013dqa,Bellazzini:2017neg}. Originally, the axion was introduced \cite{Weinberg,Wilczek} in the Peccei--Quinn solution \cite{Peccei:1977hh, Peccei:1977ur} to the strong CP problem in Quantum Chromodynamics (QCD), see e.g.\ \cite{Ringwald:2024uds} for a recent short review. 
ALPs are pseudo-Goldstone bosons which appear in two-scale models with spontaneous and explicit, but tiny, breaking of a global $U(1)$ symmetry; the axion is a prototypical example of an ALP. 
These particles are naturally light and interact with SM particles feebly. As a result, they are almost stable at cosmological time scales, so axions and ALPs are among the physically well motivated candidates for dark-matter particles \cite{Preskill:1982cy,Abbott:1982af, Dine:1982ah,ALP-DM}.

Numerous approaches have been developed to detect axions (for a review, see, e.g.  \cite{IrastorzaRedondoRev}), but have not succeeded so far. Several experimental techniques are based on the ALP interaction with the electromagnetic field,  described by the Lagrangian
\begin{equation}
\label{Lagr}
    \mathcal{L} = \frac{1}{2}(\partial_\mu a)^2 - \frac{1}{2}m_a^2\,a^2 - \frac{1}{4} g_{a\gamma\gamma}\,a\, F_{\mu\nu}\widetilde{F}^{\mu\nu}, 
\end{equation}
where $a$ is the ALP field, $m_a$ is the ALP mass, $F_{\mu\nu}$ and $\widetilde F_{\mu\nu}$ are the electromagnetic field tensor and its dual tensor, $g_{a\gamma\gamma}$ is the coupling constant, and we adopt the natural in particle physics system of units with $\hbar = c =1$. For ALPs, $g_{a\gamma\gamma}$ and $m_a$ are two independent parameters, while for ``QCD axions'', related to the solution of the strong CP problem, 
\begin{equation}
    \label{QCDaxion}
    g_{a\gamma\gamma} = 10^{-10}\, \text{GeV}^{-1}\, C_\gamma \, \frac{m_a}{0.5\, \text{eV}},
\end{equation}
where $C_\gamma$ is a model-dependent coefficient. Two particular benchmark models are those by Kim, Shifman, Vainshtein and Zakharov (KSVZ; $C_\gamma \approx 1.92$) \cite{Kim:1979if, Shifman:1979if} and by Dine, Fischler, Srednicki and Zhitnitsky (DFSZ; $C_\gamma \approx 0.75$) \cite{Dine:1981rt, Zhitnitsky:1980tq}. In what follows, we use the term ``axion'' for both QCD axions and ALPs. 

There are three large groups of approaches to searches for axions; see \cite{IrastorzaRedondoRev,Axion-astro-rev} for more details. Firstly, one can attempt both to produce and to detect axions in a controllable laboratory setup. These setups include ``light shining through walls'' and photon polarization experiments. Despite being the weakest, the constraints obtained in these experiments are the least model dependent. Secondly, it is possible to detect axions produced in natural sources, like those born in the Sun or forming the dark-matter halo of our Galaxy. These experiments test particular models of axion production, and their results are formulated under certain assumptions about the axion flux or number density. The third option is to look for indirect manifestations of axions in astrophysical environments, e.g.\ for their effects on stellar evolution, supernova energetics, cosmological observables, or propagation of high-energy photons. The latter constraints are the most abundant and often the most stringent, but they rely on certain assumptions about cosmic magnetic fields and are therefore less robust \cite{LT,PDG}.

The experiment we propose here, the Cosmological Axion Sarov Haloscope (CASH), will implement the haloscope approach, assuming that axions constitute the main component of the dark matter. We review this approach in some detail in Sec.~\ref{sec:DM}, describe the proposed experiment in Sec.~\ref{sec:CASH}, and estimate its sensitivity in Sec.~\ref{sec:sens}. Section~\ref{sec:concl} presents our brief conclusions.

\section{Dark-matter axions and haloscopes}
\label{sec:DM}
\subsection{Axion dark matter}
\label{sec:DM:DM}
Axions are among the best motivated candidates for particles forming cold dark matter, and a plethora of models have been suggested; see, e.g. \cite{Doddy-rev,Snowmass-axion-DM,PDG,OHare-axionDM-rev,Baryakhtar2025-rev} for reviews and extensive lists of references. In the early universe, axions were produced non-thermally. For a range of masses and couplings, a coherent cloud of dark-matter nonrelativistic axions can have the required energy density, that is about 
\[
\rho_a \approx 0.45\,\frac{\text{GeV}}{\text{cm}^3}
\]
in the Earth's neighborhood. 

Axions produced before inflation by means of the vacuum realignment mechanism could have very different masses, although the range of $(1-100)~\mu$eV is preferred to obtain the required abundance without fine tuning. Post-inflationary production models would be much more predictive for vacuum realignment, but complications arise because of the contribution of alternative mechanisms of axion production involving topological defects. Examples of particular models that predict post-inflationary QCD-axion dark matter with masses in the range of the CASH sensitivity can be found, e.g. in \cite{SMASH,Cogenesis,Buschmann2022Feb,VISHnu,Buschmann2024}.

\subsection{The haloscope concept}
\label{sec:haloscope}
Dark-matter axions can resonantly converse to photons in the magnetic field. This provides a way to directly detect axionic dark matter in the so-called haloscopes. Particular detection methods depend on the axion mass, which is not predicted theoretically. The idea for radio frequency mass range \cite{Sikivie:1983ip} is to construct a resonant cavity, in which the conversion is amplified by the quality factor $Q$. Due to amplification, smaller values of $g_{a\gamma\gamma}$ can be probed. The resonance condition relates the cavity eigenfrequency, and hence geometry, to the axion mass. Consequently,  to test various values of the mass, it is necessary to change the geometry of the cavity. 

\paragraph{Signal.}
The power of the electromagnetic signal, generated by the conversion of dark-matter axions in a resonant cavity filled with a magnetic field $\vec{B}_0$ is
\begin{equation}
	\label{eq:power}
	P_{a}=\left( \frac{g_{a\gamma\gamma}^2}{m_a^2}\, \rho_a \right) \times
	\bigg( F(\beta) \omega_c B_0^2 V C_{\alpha} Q_0 \bigg)\:.
\end{equation}
(see, e.g. 
\cite{Sikivie:1983ip,Krauss:1985ub,Sikivie:1985yu} and review \cite{Sikivie:2020zpn}). Here, the terms in the first brackets depend solely on the axion dark-matter parameters, while those in the second ones are determined by the characteristics of the experimental setup. The latter include the cavity volume $V$, the magnetic field $B_0$, the cavity's frequency $\omega_c=2\pi\nu_c$, the quality factor $Q_0$, and the factor $F(\beta)=\beta/\left(1+\beta\right)^2$, where $\beta$ is the coupling between the cavity and a receiver. The form factor $C_{\alpha}$ \cite{Sikivie:2020zpn} reads
\begin{equation}
    C_\alpha \equiv \frac{1}{B_0^2}\int_V \frac{d^3 x}{V}\left( \vec{B}_0 \cdot \vec{e}_\alpha (x)\right)^2,
\end{equation}
where $\vec{e}_\alpha$ is the electric field for the cavity mode $\alpha$, normalized to the Kronecker tensor $\delta_{\alpha \beta}$ as  $\int_V d^3 x\left(\vec{e}_\alpha (x) \cdot \vec{e}_\beta (x)\right)=\delta_{\alpha \beta}$,  and the magnetic field is assumed to be uniform across the cavity volume. The cavity is considered invariant under translation in the direction of the magnetic field $\vec{B}_0$, say the $\hat{z}$ axis, except for the cavity end caps. Only transverse magnetic (TM) cavity modes yield a non-zero form factor. In the case of the cylindrical cavity, the lowest mode is TM$_{010}$, and the corresponding form factor is $C_\alpha=C_{010}=0.69$. The cavity is sensitive to axion masses $m_a = \omega_c$ with a relative accuracy of about $1/Q_L$, where the loaded quality factor $Q_L\equiv Q_0/(1+\beta)$.

The function $F(\beta)$ reaches its maximal value of 0.25 at $\beta=1$, which corresponds to the critical coupling of a cavity and a detector. Since achieving the critical coupling is mainly an engineering problem, for estimates, we substitute in eq.~(\ref{eq:power}) $F(\beta)=0.25$, and  the unloaded quality factor $Q_0=10^4-10^5$ for a copper cavity. Let us consider a simple cylindrical cavity with a tuning rod as in \cite{Semer_1989}, which allows for 10\% to 20\% frequency tuning. Then, even the maximal possible tuning of the frequency in this cavity design leads to a rather weak variation of the coupling function $F(\beta)>0.2$, but, as we show below, may significantly affect the form factor $C_{010}$, making it much smaller than 0.69.

At low values of the axion power (\ref{eq:power}), the axion conversion should be described in terms of the generation of single electromagnetic quanta (photons). Since the initial axion field in a cloud is highly coherent, the photons generated in the cavity obey the Poisson distribution \cite{fox2006quantum}.

\paragraph{Noise.}

In experiments with cavity-based haloscopes, there are two main sources of noise. Firstly, there exist thermal photons, generated in the cavity because of a finite temperature. The second source is the detector noise, which may appear due to various reasons, such as internal thermal and quantum fluctuations and various background radiation sources.

For a given cavity frequency $\omega$, the thermal photon flux obeys the Planck distribution, so at low temperatures $T$ this source of noise is exponentially suppressed $P_{\rm th~phot} \propto \exp(-\omega/T)$. For $\omega = 10$ GHz, this critical temperature is $T \sim 30$~mK. The lower temperature below $20$~mK is easily achievable in modern dilution refrigerators, allowing the flux of thermal photons to be suppressed. 
Unlike coherent photons, thermal ones emitted from a single mode of a cavity obey the Bose-Einstein (super-Poisson), rather than the Poisson distribution \cite{fox2006quantum,Pankratov2025Apr}.

If the detector noise (e.g. dark counts) dominates over other sources of noise, then the statistical analysis can be performed with the standard signal-to-noise ratio (SNR) formalism as in \cite{Graham:2023sow}, 
\begin{equation}
\label{eq:SNR}
    \mbox{SNR} = \frac{N_{\rm sig}}{\sqrt{N_{\rm sig}+N_{\rm d.c.}}} = \frac{R_{\rm sig}}{\sqrt{R_{\rm sig}+R_{\rm d.c.}}}\,\sqrt{t},
\end{equation}
where $t$ is the operation time, $N_{\rm sig\, (d.c.)}$ is the number of signal (dark count) events during the operation time, and $R_{\rm sig\,(d.c.)} = N_{\rm sig\, (d.c.)}/t$ is the signal rate,
cf. \cite{Graham:2023sow}.  In the case of dominating thermal photons, not described by the Poisson statistics \cite{Pankratov2025Apr}, Eq.~(\ref{eq:SNR}) should be corrected.

\subsection{Landscape of experimental efforts}
\label{sec:DM:others}

Most of the past and present radio-frequency cavity haloscope experiments have used microwave detectors, constrained by the Standard Quantum Limit (SQL). These include classical heterodyne receivers or quantum microstrip-coupled SQUID amplifiers \cite{Prokopenko2003Jun,Spietz2008Aug}, as well as Josephson parametric amplifiers (JPAs) \cite{Castel_2008,Fedorov_2019,Qiu_2023}. The latter are also limited at the level close to SQL,  which in terms of noise temperature reads (in SI notations)
\begin{equation}
\frac{T_{SQL}}{100\,\text{mK}}=\frac{h}{k_{B}}\frac{f}{2\,\text{GHz}}.   
\end{equation}
It was shown that SQL can be slightly overcome using squeezed states in JPAs \cite{Castel_2008,Fedorov_2019,Qiu_2023, Malnou:2018dxn}. 

The most sensitive cavity haloscope experiments (see Table~7 in Ref.~\cite{Antel:2023hkf} for a more detailed list) in the axion mass range $2 - 10 \ \mu$eV are the Axion Dark Matter Experiment (ADMX) \cite{ADMX_proj,ADMX1,ADMX2,ADMX3} and projects in the Center for Axion and Precision Physics Research (CAPP) \cite{CAPP1,CAPP2,CAPP3}. Smaller axion frequencies correspond to larger cavities and lead to less stringent SQL restriction, so, at higher magnetic fields, the use of linear amplifiers as detectors may lead to the desired DFSZ sensitivity. Cavity experiments aimed at testing axion models with larger masses,  $(20-100)~\mu$eV, include the QUaerere AXions (QUAX) \cite{QUAX-1,QUAX-2} 
and the Haloscope At Yale Sensitive To Axion Cold Dark Matter (HAYSTAC)  \cite{HAYSTAC:2018rwy,HAYSTAC:2023cam} experiments. In the second version of HAYSTAC, squeezed state detectors are used \cite{HAYSTAC:2023cam}.

Among alternatives to the microwave cavities are broadband dish antenna haloscopes \cite{Horns:2012jf}, aimed at higher frequency range above 50 GHz, where a cavity concept is difficult to implement due to smaller sizes even in multi-cavity configurations. The corresponding experimental setups include Broadband Radiometric Axion SearcheS (BRASS) \cite{Bajjali:2023uis} and Broadband Reflector Experiment for Axion Detection (BREAD) \cite{Hoshino:2025fiz} projects. The same principle as for the dish antenna works for the dielectric disk experiment, e.g. the Magnetized Disc and Mirror Axion experiment (MADMAX) \cite{MADMAX_proj,MADMAX}. There is also the concept of a plasma haloscope, where the plasma frequency is tuned to the axion mass. Technically, it can be achieved by exploiting metamaterials as in the Axion Longitudinal Plasma HAloscope (ALPHA) project \cite{ALPHA_proj, ALPHA:2022rxj}. 

\section{The CASH experiment}
\label{sec:CASH}
\subsection{Single-photon detectors}
\label{sec:CASH:detectors}
It is currently understood \cite{Lamoreaux2013Aug,Sushkov2023May} that single-photon detectors (SPDs) are of utmost importance for axion search in the frequency range above 1~GHz. Microwave SPDs based on Josephson junctions \cite{Pankratov2025Apr,Pankratov2022May,Ladeynov_2025} and superconducting qubits \cite{Lescanne2020May,Albertinale2021,Balembois2024Jan} are becoming available, and below we focus on their use in the proposed project. 

\begin{figure}[h!]
    \centering
    \includegraphics[width=\linewidth]{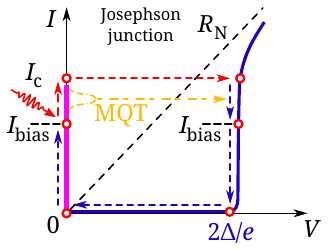}
    \caption{Sketch of the hysteretic current-voltage characteristic of an underdamped Josephson junction as a single-photon detector. Red dashed lines -- switching due to a photon, blue dashed lines -- restoring of the initial state. MQT means false switching due to either tunneling or thermal escape.}
    \label{fig:IVC}
\end{figure}
The underdamped Josephson junctions (JJs) with hysteretic current-voltage characteristics \cite{Pankratov2022May,Pankratov2025Apr,Ladeynov_2025} are planned to be used in the CASH experiment; see Fig.~\ref{fig:IVC}. 
Such SPDs represent threshold detectors operating in the superconductive state. We supply a bias current $I_{\text{bias}}$ close to the critical current $I_\text{c}$. Due to photon arrival, the JJ switches to the resistive state with the gap voltage $V=2\Delta/e\sim 0.4\, $mV for Al, with $\Delta$ being the superconductive gap and $e$ referring to the electron charge. To return to the working state, the current through JJ should be set to zero and after that should be increased back to $I_{\text{bias}}$. The response time of a detector is a fraction of ns, while the dead time of the detector, needed to restore the initial state, is about $1\, $ms, being restricted by the used $RC$-filters.
JJs are implemented into coplanar quarter-wavelength resonators with working bandwidth about 10\% of the central frequency. The JJ samples are fabricated as Al-Al\,O$_x$-Al trilayers using self-aligned shadow evaporation technique without breaking the vacuum. First, contact pads, direct-current lines, and coplanar resonator layer are fabricated using laser lithography and electron beam evaporation. After that, electron lithography and shadow evaporation are performed for the deposition of the first aluminum layer, its oxidation in a load-lock chamber, and the deposition of a thicker final aluminum layer. While some of the JJs can detect photons in several frequency ranges, it is necessary to vary coplanar resonators to match the frequency ranges of copper cavities, so many JJ SPD samples are required.

Since the flux of photons, appearing due to axion conversion in a high magnetic field (signal), is expected to be rather low, the dark count rate of a detector (noise) should be made as low as possible. 
For single-photon detectors, in addition to the flux of thermal photons from the cavity, there is a non-thermal component, leading to false switching of the detector, the so-called dark counts. These dark counts for the Josephson-junction-based SPDs obey the Poisson statistics \cite{Pankratov2025Apr}. Dark counts come from various sources, e.g. internal erroneous switches due to thermal activation or quantum tunneling; low-frequency interference due to compressors and pumps present in the setup; cosmic rays or other sources of radiation; and electromagnetic interference due to various sources. Among these sources is the stray light from the higher cryostat plates, which operate at higher temperatures \cite{Barends2011Sep}. This motivates placing the detector on the coldest plate, surrounding it with a double shield \cite{Barends2011Sep}. All these sources, leading to false switches of a detector, are measured directly as a basic qualification of SPD dark counts \cite{Pankratov2025Apr} with the dark count rate reaching 0.01 Hz, with possible improvement to 0.001 Hz and below.

Although presently only a dark count rate of about 0.01~Hz has been demonstrated, it was limited by the time it took to collect reliable statistics only. Based on the available data for the sample, analyzed in \cite{Pankratov2025Apr}, lower dark count rates can be achieved, see below. 

\begin{figure}[h!]
    \centering
    \includegraphics[width=\linewidth]{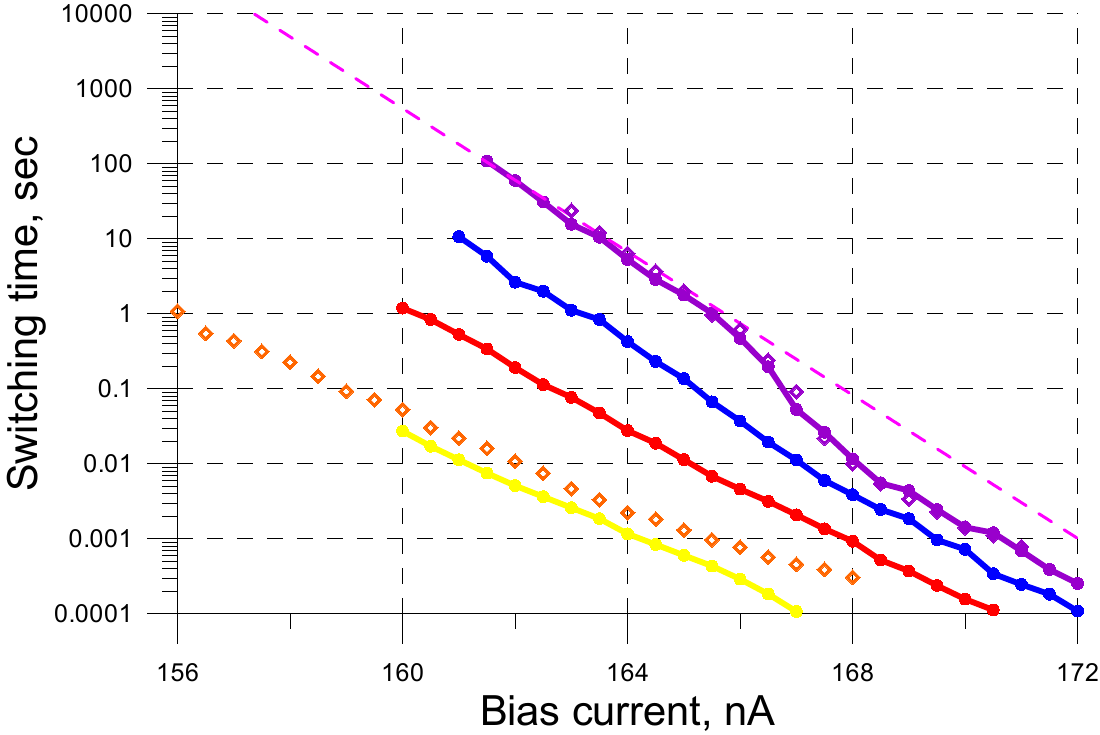}
    \caption{Plot of the mean switching time of a single photon detector, showing feasibility of dark count time above 100~s (violet dots and magenta dashed curve). Other curves correspond to heating the copper cavity from 21 mK to 40, 60 and 80 mK (blue, red, yellow/orange dots), respectively.}
    \label{fig:LT}
\end{figure}
The plot of the mean switching time of a single-photon detector is presented in Fig.\,\ref{fig:LT} for various temperatures of the copper cavity used as a source of thermal photons. In the dark count regime, i.e. at the base temperature of the photon source of 21~mK (violet symbols at the top), the mean switching time is the inverse of the dark count rate and exceeds 100 s, corresponding to $\sim 0.01$ Hz of the rate. Other curves correspond to a mixture of dark counts and switching due to the arrival of thermal photons from the cavity, heated to 40, 60 and 80 mK, thus leading to a significant reduction of the switching time. The results of two different measurement cycles are presented. In the first cycle (circles, connected by solid curves in Fig. \ref{fig:LT}), the bias current range is restricted by 160~nA only. One can fit the corresponding dots by a magenta dashed line and see that at 160~nA, the dark count time value approaching 600~s can be reached. In the other cool-down, the switching time curve for $\approx 80$~mK cavity temperature has been measured to 156~nA (orange diamonds). One can see that this curve roughly coincides with the corresponding yellow curve and keeps the same shape without bending up, with a certain reduction of switching efficiency. By prolonging the dashed magenta curve further, one can see that the dark count time can exceed $10^4$~s, while the ratio between the dashed magenta and orange data points exceeds $10^4$. The feasibility of reaching so large dark count times depends only on the measured system stability, which can be improved by proper filtering and damping systems to decrease vibrations and microphonic effects due to pumps and compressors of the cryostat. Actually, values above 1000~s have been reached for samples with smaller critical currents \cite{Pankratov2022May}.
\begin{figure*}
    \centering
    \subfigure(a){\includegraphics[height=127pt]{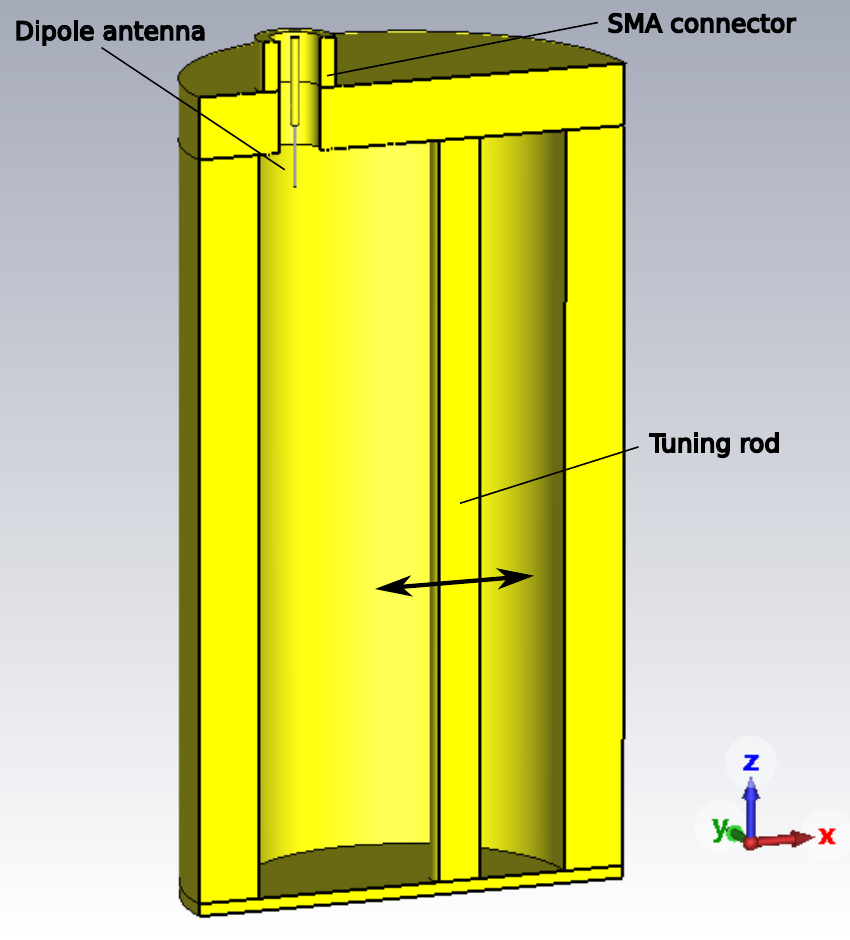}}
    \subfigure(b){\includegraphics[height=127pt]{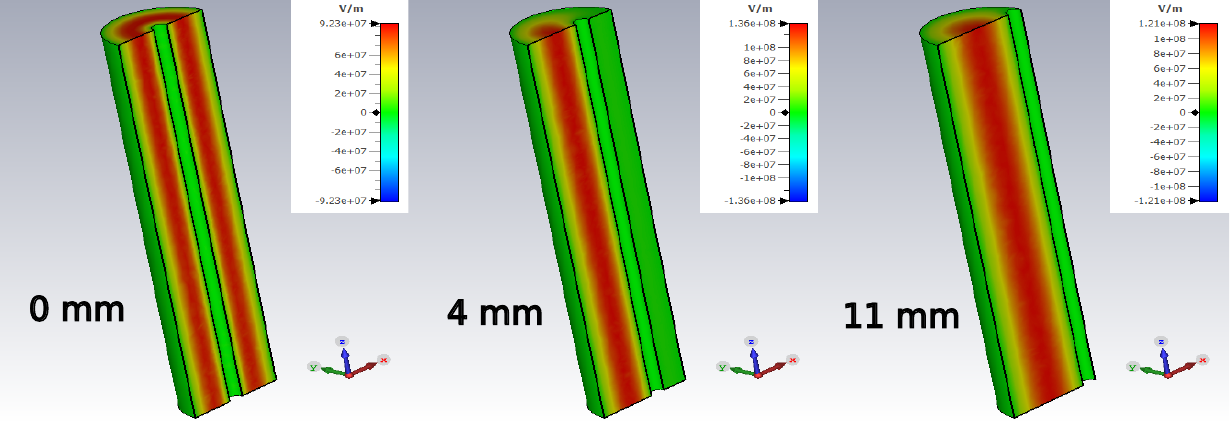}}
    \caption{(a) A cut view of the cavity with a  SMA connector. The cavity has inner height of 120 mm and inner diameter of 26 mm. A metallic tuning rod with 4 mm diameter is used to tune the cavity to a desired frequency by moving it along the radius of the cavity. (b) A structure of the uniform TM mode of the cavity at the rod shift equal to 0, 4, and 11 mm.}
    \label{fig:cavity}
\end{figure*}

\subsection{Cryostat, magnet and cavity}
\label{sec:CASH:setup}

For measurements in the initial stage, it is planned to use the dilution refrigerator with base temperature $\sim10$~mK and the cooling power of 200~$\mu$W at 100~mK.
The duration of a cryogenic experiment with 20~mK base temperature of a cavity can be as long as three or four weeks. In some cases, for longer runs, the base temperature can slightly rise, which should be checked in additional experiments. The available $1.7$~T magnet is intended to be placed at 1~K thermal screen. For the second stage of the CASH experiment, a 7~T or 10~T magnet is planned to be placed at  4~K thermal screen.

In the initial stage, we plan to use a cylindrical copper cavity with a length of 120 mm and a diameter of 26 mm, which corresponds to a cavity volume $V = 63.7 $ $\mbox{cm}^3$. The cavity-detector coupling is taken to be critical with $\beta=1$, corresponding to $F(\beta)=0.25$.

As the axion mass is unknown, the haloscope cavity eigenfrequency should be tuned. Frequency tuning within a bandwidth of $(10 - 30) \%$ can be carried out by insertion of moving dielectric \cite{Semer_1989} or metal \cite{ADMX3} rods or shells into the cavity. The frequency tuning with moving parts is reviewed in \cite{Bai_2023}.
Dielectric insertions reduce the cavity axion form factor because of localization of a mode field inside the dielectric. Our simulations performed with the help of the CST Studio Suite microwave simulation package reveal that a sapphire rod of $4$ mm diameter, pushed by $10$ mm into the cavity of $120$ mm length, decreases the eigenfrequency of the TM010-like mode by $14\%$. Simultaneously, the form factor $C_\alpha$  decreases dramatically from $0.693$ to $0.107$.  Thus, following ADMX \cite{ADMX3}, HAYSTAC \cite{HAYSTAC:2023cam}, QUAX \cite{QUAX-2}, CAPP \cite{CAPP-tune} and other haloscope experiments, we prefer to tune the cavity with a copper rod of the same 4~mm diameter moving it along the radius of the cavity, see Fig.~\ref{fig:cavity}a.

\begin{figure}
    \centering
    \subfigure(a){\includegraphics[width=\linewidth]{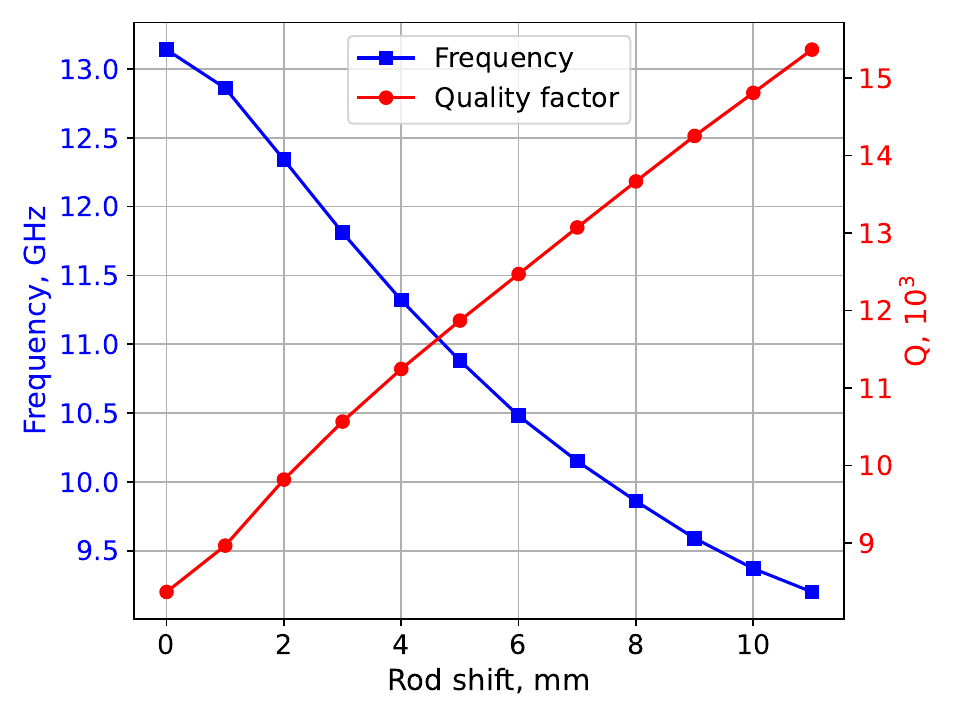}}
    \subfigure(b){\includegraphics[width=0.9\linewidth]{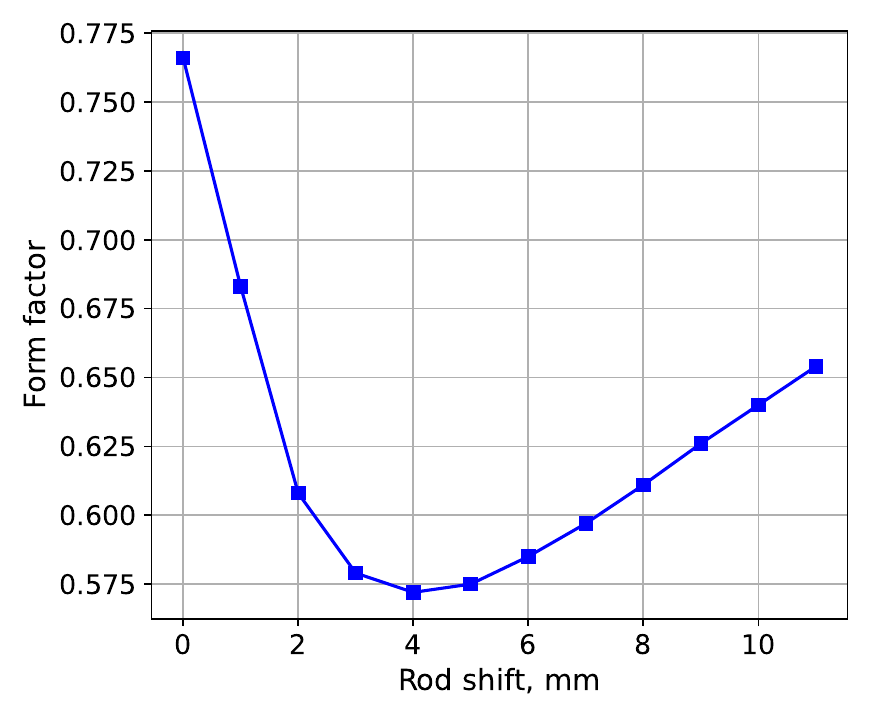}}
    \caption{The dependence of (a) resonance frequency $\omega_c$ (blue curve) and quality factor $Q_0$ (red curve) and (b) geometric form factor $C_{\alpha}$ on the shift of the metallic rod from its on-axis position. }
    \label{fig:fqformfactor}
\end{figure}

We perform numerical simulations to show the distribution of the electric field of the cavity TM010-like mode with the uniform structure along the cavity axis, for a set of positions of the copper rod, see Fig.~\ref{fig:cavity}b. This mode is expected to have the largest form factor among all. Further, we have calculated the resonance frequency $\omega_c$, the quality factor $Q$, and the form factor $C_\alpha$ for the aforementioned TM010-like mode as a function of the shift of the tuning rod from its on-axis position; see Fig.~\ref{fig:fqformfactor}.
The range of the rod shift is from 0~mm (on-axis position) to 11~mm, which corresponds to the configuration where the rod touches the cavity wall. In Fig.~\ref{fig:fqformfactor}a, we see that the resonance frequency decreases almost linearly from 13.14~GHz to 9.2~GHz while the rod shifts. Likewise, the quality factor $Q_0$ grows almost linearly from $8.3\times10^3$ to $15.6\times10^3$. Note that the $Q_0$ dependence is calculated for copper at room temperature and will increase at lower temperatures. Fig.~\ref{fig:fqformfactor}b shows that the form factor $C_\alpha$ at first drops rather sharply from the initial value of 0.766 to the minimum value of 0.572, reached as the rod shifted by 4 mm. Then, while the rod is shifted further, the form factor gradually grows and reaches 0.654 with the rod touching the wall. We see that even in the minimum, the form factor takes a rather large value.

There is an important question about the speed of tuning the cavity frequency. Using a piezoelectric motor, placed on a higher plate of a cryostat, one can perform rather precise and fast frequency tuning by moving the rod inside the cavity. Here, to control the achieved frequency band, one can exploit an external synthesizer to supply a tone signal, while the same JJ SPD can be used to measure the cavity frequency band. 

\begin{figure}
    \centering
    \includegraphics[width=\linewidth]{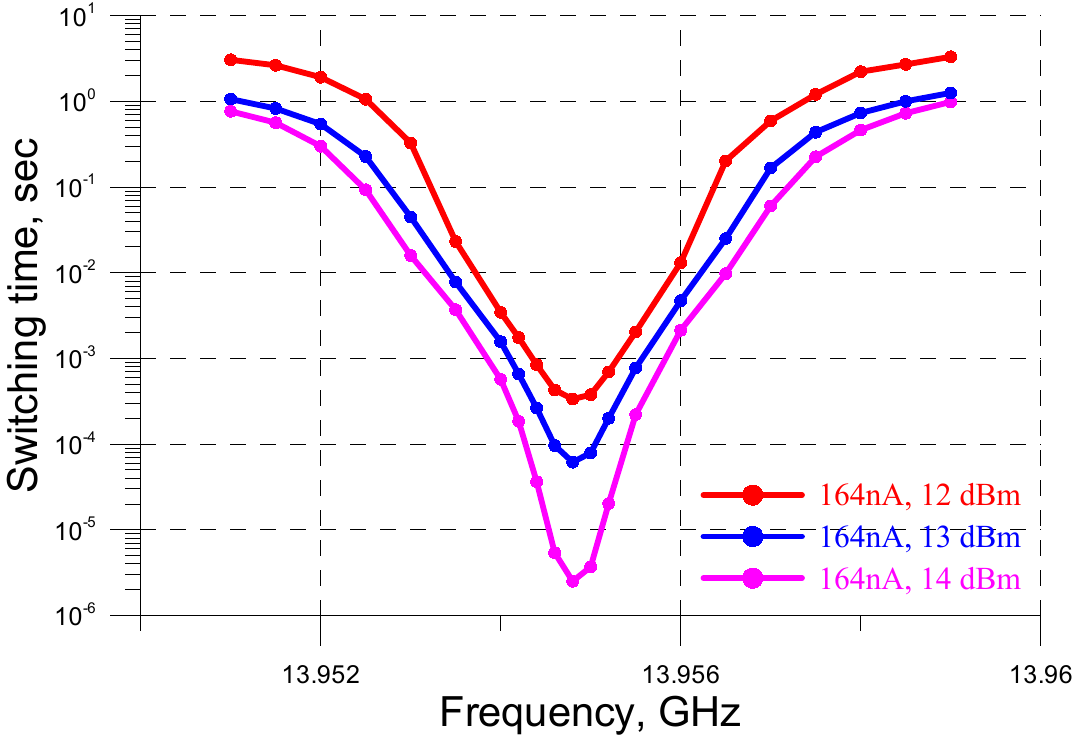}
    \caption{Measuring cavity bandwidth using a JJ SPD.}
    \label{fig:cavityBand}
\end{figure}

Since calibrating cavity resonances using Photon-Assisted Tunneling (PAT) steps at the inverse branch of the current-voltage characteristic \cite{Pankratov2025Apr} is rather time-consuming, one can perform such calibration once and later measure the frequency dependence of the detector switching time by a fast sweep. An example of switching times, reflecting the cavity frequency band, is shown in Fig. \ref{fig:cavityBand} 
for various powers of a test signal (attenuated at cold stages by about 60 dB). This approach significantly outperforms the use of a vector network analyzer (VNA) for the following reasons. First, with JJ SPD the only one coaxial line is needed to supply a test tone to the cavity, see Fig.\,\ref{fig:cavity}a. Therefore, a cavity should only have two antenna ports: one critically coupled to be connected with a detector and another weakly coupled to supply test signals. This second port only slightly affects the quality factor of the cavity. Second, one needs to protect both the cavity and the detector from the room-temperature background. In this case, one should use about 60~dB of cold attenuators, as for qubit readout. In case of using a room temperature VNA to control the frequency tuning, at least two coaxial lines are needed, one of which is supplied with a cold amplifier; otherwise, the VNA sensitivity will not be sufficient. 

\section{CASH sensitivity}
\label{sec:sens}
The signal rate (mean number of signal events per second) reads $R_{\text{sig}} = P_a / m_a$, where the signal power $P_a$ is determined by  eq.~(\ref{eq:power}), precisely
\begin{equation}
	\label{eq:rate}
	R_{\text{sig}}= C_\gamma^2 \; B_0^2 \; Q_0 \times G.
\end{equation}
Here $C_\gamma$ is determined by eq.~\eqref{QCDaxion}, and
\begin{equation}
\label{eq:G}
    G \equiv \rho_a V F(\beta)C_{\alpha}\times 0.04 \ \mbox{GeV}^{-4}. 
\end{equation}

To derive the setup sensitivity for a given axion model, we use the SNR approach following Eq.~(\ref{eq:SNR}). 
The one-sided $95 \%$ CL exclusion level corresponds to $\rm{SNR}  = 1.65$, while $\rm{SNR} = 5$ corresponds to the traditional definition of a ``discovery''. 

The measurement time $t$ required to exclude the given benchmark axion model (fixed $C_\gamma$) at a given CL reads  
\begin{equation}
    t = \frac{R_{\rm sig}+R_{\rm d.c.}}{R_{\rm sig}^2}\times \mbox{SNR}^2.
\end{equation}
It is instructive to fix the required observation time $t$, and obtain the sensitivity for $C_\gamma$,
\begin{align}
\label{eq:sensitivity}
C_\gamma = \frac{1}{B_0 \sqrt{G\, Q_0}}  \times \frac{\mbox{SNR}}{\sqrt{2t}} \left(  1+ \sqrt{1 + 4\frac{t \, R_{\text{d.c.}}}{\mbox{SNR}^2}}\right)^{1/2}.
\end{align}
For given setup parameters and fixed exclusion level (SNR sigma), one would exclude all values of $C_\gamma$ exceeding that determined by Eq.~(\ref{eq:sensitivity}).

\subsection{CASH I: fixed frequency}
\label{sec:sens:I}
CASH I is the first stage of the proposed experiment, at which we plan to use a non-tunable cavity, thus the resonant frequency will be fixed at a single value. This stage uses a magnet with $B_0 = 1.7$~T described above; the cavity is a copper cylinder of length of 120~mm and a diameter of 26~mm, for which $V=63.7\,\mbox{cm}^3$. Benchmark values are taken for the TM010 mode eigenfrequency $\omega_c = 2\pi \cdot 8.8$ GHz, corresponding to $m_a = 36.5 \,\mu\mbox{eV}$, assuming the unloaded quality factor at mK temperatures $Q_0 = 10^5$ and the form factor $C_\alpha=C_{010}=0.69$. Equation~(\ref{eq:power}) gives the expected signal photon rates: 0.012~Hz for KSVZ model and 0.0019~Hz for DFSZ model, assuming critical coupling of the cavity and the detector, as stated above. 
A possible future improvement is to use a stronger field magnet. With a $7$~T magnet of the same geometric dimensions and the same cavity design, the expected KSVZ and DFSZ signal photon rates would reach $0.2$~Hz and $0.03$~Hz, respectively.

\begin{figure}
    \centering    \includegraphics[width=0.999\linewidth]{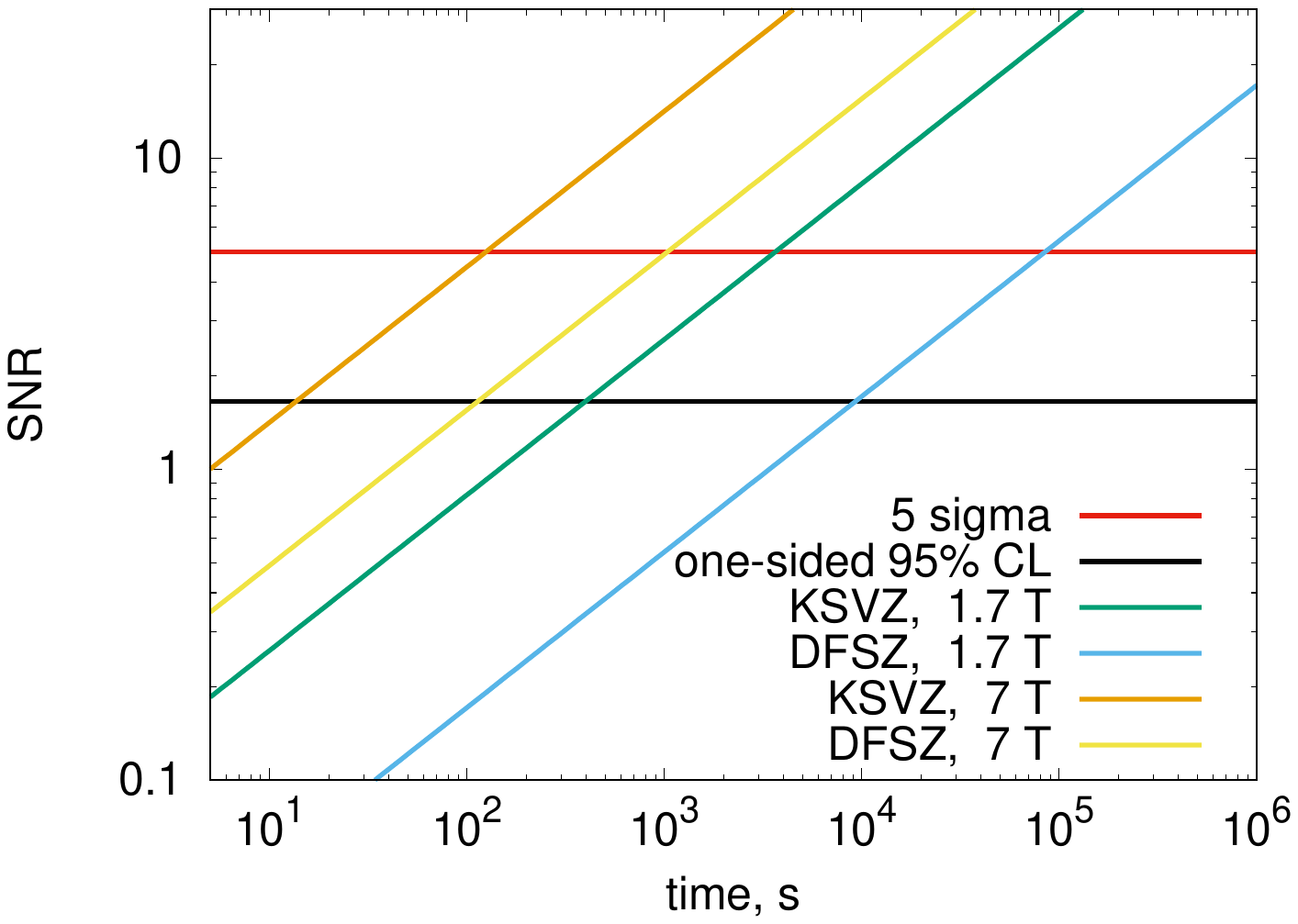}
    \caption{Signal-to-noise ratio for two benchmark models (KSVZ and DFSZ) for two values of magnetic field $B=1.7$\,T, $7$\,T; $Q_0=10^5$. SNR$=5$ means the $5\sigma$ discovery of the given benchmark axion model for a fixed mass, SNR$=1.65$ is related to one-sided $95\%$ CL exclusion.}
    \label{fig:enter-label}
\end{figure}

Let us determine the observation time required to test KSVZ (DFSZ) models for different values of the magnetic field. The SNR dependence on time is shown in Fig.~\ref{fig:enter-label}. 
Thus, $t=400\, (9200)$ s, or $6.5$ minutes ($2.5$ hours) are needed to probe KSVZ (DFSZ) with the magnetic field of $1.7$\,T. For the stronger magnet of 7\,T, the time reduces to 14~s for KSVZ or 112~s for DFSZ, see Table~\ref{tab:obs_time_1}.
\begin{table}
    \centering
    \begin{tabular}{|c|c|c|}
    \hline
      & $B$, T  &  $t$  \\ \hline 
    KSVZ & 1.7     &  $400\,\mbox{s} \simeq 6.5 \, \mbox{min}$  \\ 
    DFSZ & 1.7     & $9\,200\,\mbox{s} \simeq 2.5 \, \mbox{hours} $  \\
    KSVZ & 7     &  $14 \,\mbox{s}$ \\
    DFSZ & 7     &  $112 \,\mbox{s}$  \\ \hline
    \end{tabular}
    \caption{
    Observational time required to reach the 95\% CL axion model sensitivity for $Q_0=10^5$.}
    \label{tab:obs_time_1}
\end{table}
With extended observation time, we may probe the coupling $C_\gamma$ beyond the DFSZ prediction. 
Let us use the formula (\ref{eq:sensitivity}) for sensitivity. Substituting additionally $R_{d.c.} = 0.01$~Hz, $t = 10^6$~s (12 days) and SNR $=1.65$ (one-sided $95\%$ CL exclusion), one obtains
\begin{equation}
\label{eq:Ninoax_g}
 C_\gamma = 0.22, \quad \text{or} \quad  g_{a\gamma\gamma} = 1.6 \cdot 10^{-15} \;\mbox{GeV}^{-1},
\end{equation}
where we assume 1.7~T magnetic field and the axion mass $m_a=38\,\mu$eV fixed for the CASH~I geometry. Thus, all theoretically favored models of the Peccei-Quinn axion could be fully explored for this mass. 
\begin{figure}
    \centering
    \includegraphics[width=\linewidth]{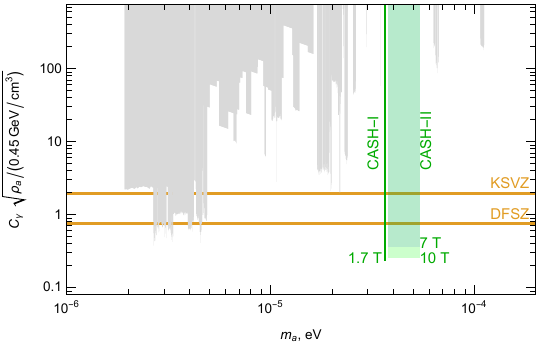}
    \caption{%
    Projected sensitivity of the first (CASH-I; non-tunable, $10^6$~s of observation, $1.7$~T), and second (CASH-II, tunable,  1~year of observation, magnetic field $7$ or $10$ T) stages of the proposed experiment for the axion-photon coupling coefficient $C_\gamma$, together with the regions excluded by Refs.~\cite{ADMX1,ADMX2,ADMX3,ADMX4,ADMX5,ADMX6,ADMX7,RBF1,RBF2,UF1,UF2,CAPP1,CAPP2,CAPP3,CAPP4a,CAPP4b,CAPP5,CAPP6,CAPP7,CAPP8,CAPP-MAX,CAST-CAPP,HAYSTAC:2023cam,TASEH,GrAHal,ORGAN1,ORGAN1a,ORGAN1b,ORGAN-Q,RADES-1,RADES-2,QUAX-1,QUAX-2,QUAX-3,QUAX4,QUAX5,MADMAX}, mostly compiled in \cite{AxionLimits} (gray shading). All limits are rescaled to the 95\% CL assuming Gaussian statistics. Values of $C_\gamma$ for two benchmark modes of the QCD axion are indicated as horizontal lines. }
    \label{fig:sensitivity1}
\end{figure}
\begin{figure}
    \centering
    \includegraphics[width=\linewidth]{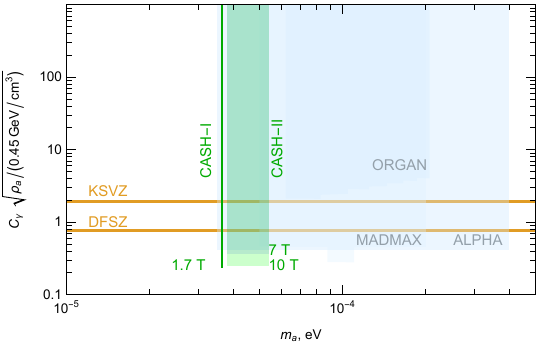}
    \caption{%
   Projected sensitivity of the first (CASH-I; non-tunable, $10^6$~s of observation, $1.7$~T), and second (CASH-II, tunable,  1~year of observation, magnetic field $7$ or $10$ T) stages of the proposed experiment for the axion-photon coupling coefficient $C_\gamma$,  together with projected sensitivities of ALPHA \cite{ALPHA_proj,ALPHA:2022rxj}, ORGAN \cite{ORGAN_proj}, and MADMAX \cite{MADMAX_proj}. All limits are rescaled to the 95\% CL assuming Gaussian statistics. Values of $C_\gamma$ for two benchmark modes of the QCD axion are indicated as horizontal lines. 
    }
    \label{fig:sensitivity2}
\end{figure}

Even for the magnetic field $B_0=1.7$~T, the sensitivity of CASH~I is unprecedented due to its very low noise level and the use of single-photon detectors. The sensitivity (\ref{eq:Ninoax_g}) is presented in Fig.~\ref{fig:sensitivity1} as a thin green vertical line marked ``CASH-I,'' and compared to other projects in Fig.~\ref{fig:sensitivity2}.
Substituting $B_0 = 7$ T, one obtains the sensitivity of $95\,\%$ CL upper limit,
\begin{equation}
\label{eq:Ninoax_g10}
   C_\gamma = 0.05, \quad \text{or} \qquad g_{a\gamma\gamma} = 3.9 \cdot 10^{-16} \;\mbox{GeV}.
\end{equation}
The projected bounds \eqref{eq:Ninoax_g}, \eqref{eq:Ninoax_g10} are related to the single frequency of the resonant cavity, which corresponds to a very narrow line at the axion parameter plot with the width  $\delta m_a = 2 m_a/Q_0 \approx 1.9\times 10^{-10}$~eV.

\subsection{CASH II: tunable frequency}
\label{sec:sens:2and3}
\paragraph{Frequency tuning.}
In the second stage, CASH~II, we plan to scan the axion parameter space by tuning the cavity resonant frequency. Precise tuning will be performed by a tiny shift of a copper rod inside the cavity, as described in Sec.~\ref{sec:CASH:setup}. The copper rod is planned to be installed in the cavity used in CASH-I. Due to the presence of a copper rod of $4$ mm diameter, the volume of the cavity decreases slightly to $V=62.2\,\mbox{cm}^3$. The tuning would make it possible to probe axion masses in the range\footnote{The minimal axion mass shifted by $1.5\, \mu$eV compared to CASH-I due to the presence of a copper rod.} $(38- 54)\,\mu$eV with a single cavity. JJ SPDs have a narrower bandwidth $\sim 10\%$ and are located outside the cavity at the coldest 10~mK plate, connected via coaxial cable. Thus, we can replace the SPD with the one that is more sensitive to the current axion mass range several times. We assume the time of a single frequency shift as $\delta t = 15$~min, including frequency control. The dependences of $\omega_c$, $Q_0$ and $C_\alpha$ on the copper rod location are shown in Fig.~\ref{fig:fqformfactor}. However, for rough estimates, we keep the form factor $C_\alpha$ and the quality factor $Q_0$ to be constant.

Let us obtain the number $N$ of frequency shifts required to scan the available range. The width of a resonant line for a fixed frequency $m_a$ is $\delta m_a =  m_a/Q_L$, and for the critical coupling $\beta=1$ one has $Q_L=Q_0/2$. The mass interval reads $(m_a, \,m_a(1+Q_L^{-1})))$. The single tuning shifts the interval to $(m_a(1+Q_L^{-1}), \,m_a(1+Q_L^{-1})^2)$ etc. Finally, 
\begin{equation}
    m_R = m_L\left( 1+ Q_L^{-1}\right)^N,  
\end{equation}
where $m_L$ and $m_R$ are left and right bounds of the scanning interval of axion masses, $38$ and $54$ $\mu$eV in our case. 
Taking $Q_L^{-1} \ll 1$, one obtains 
\begin{equation}
    N = Q_L \log(m_R/m_L) = \frac{\log1.4}{2}\cdot Q_0.
\end{equation}
The total measurement time is $\tau = N(t +\delta t)$, where $t$ is the measurement time for a fixed frequency. The measurement time for a fixed frequency reads 
\begin{equation}
\label{eq:time}
    t = \frac{\tau}{N} - \delta t = \frac{2\tau}{Q_0 \log 1.4} - \delta t. 
\end{equation}

\paragraph{Optimization of the unloaded quality factor $Q_0$.}
The dependence of the setup sensitivity on the quality factor $Q_0$ is two-fold. On the one hand, at a fixed frequency, the sensitivity improves with increasing $Q_0$. On the other hand, the bandwidth for a fixed frequency narrows with increasing $Q_0$, so the number of frequency shifts $N$ increases, and the measurement time at a fixed frequency decreases. The sensitivity also decreases. These two opposite behaviors are represented by the analytical formula, cf. Eqs.~\eqref{eq:SNR}, \eqref{eq:rate}, \eqref{eq:time},
\begin{equation}
\label{eq:SNRQ}
 \mbox{SNR} =\! \frac{C_\gamma^2\; B_0^2 G Q_0}{\sqrt{C_\gamma^2\;B_0^2 G Q_0\!+\!R_{\rm d.c.}\!}}\sqrt{\frac{2\tau}{Q_0 \log 1.4} - \delta t} .  \!
\end{equation}
Here, $G =  0.19\ \mbox{GeV}^{-3}$, cf. Eq.~\eqref{eq:G}.

\begin{figure}
    \centering
    \includegraphics[width=0.99\linewidth]{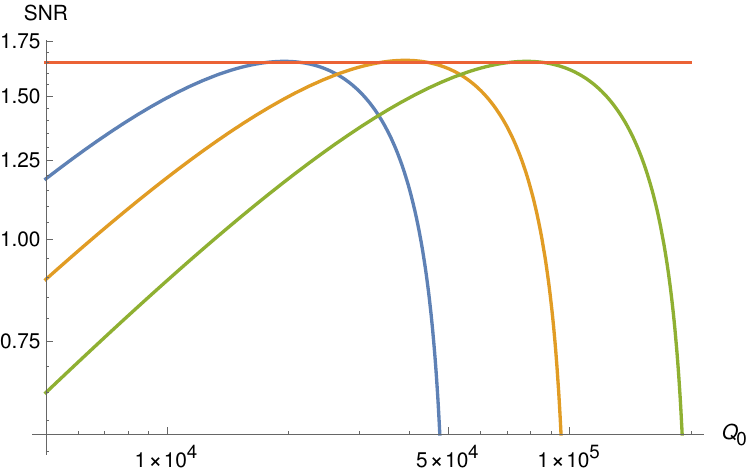}
    \caption{SNR dependence on $Q_0$ for scanning the mass range $(38 - 54)\; \mu$eV, see eq.~(\ref{eq:SNRQ}). The product $C_\gamma B_0$ is taken to obtain the maximal value SNR$=1.65$; $\delta t = 15$~min;  $R_{\rm d.c.}=0.01$~Hz; $C_\alpha = C_{010}=0.69$.  
    Left curve: $\tau=3$  months. The best sensitivity is for $Q_0=2\cdot 10^4$, $C_\gamma B_0 = 5$~T.  
    Middle curve: $\tau=6$  months. The best sensitivity is for $Q_0=4\cdot 10^4$, $C_\gamma B_0 = 3.6$~T. 
    Right curve:  $\tau = 1$ year.  The best sensitivity is for $Q_0=8\cdot 10^4$, $C_\gamma B_0 = 2.5$~T. 
    }
    \label{fig:Q}
\end{figure}

The SNR dependence on $Q_0$, Eq.~(\ref{eq:SNRQ}), illustrated in Fig.~\ref{fig:Q}, 
exhibits a maximum in the interval $Q_0 \sim (10^4 - 10^5)$. We choose the combination of parameters $(C_\gamma B_0)$ in order to obtain the SNR$=1.65$ at the maximum, for 3 different values of measurement time: 3 months, 6 months and 1 year. The corresponding optimal values of $Q_0$ are presented in Tab.~\ref{tab:tab2}.    
\begin{table}
    \centering
    \begin{tabular}{|c|c|c|c|} 
    \hline 
       total time $\tau$, years  &  0.25 & 0.5 & 1 \\ \hline
       optimal $Q_0$  &$2\times 10^4$ & $4 \times 10^4$ & $8\times 10^4$ \\ \hline
       $N$ & $3.5 \times 10^3$ & $7 \times 10^3$ & $1.4 \times 10^4$ \\ \hline
         $C_\gamma$ for $B_0=7$~T & $0.72$ & $0.51$ & $0.36$ \\ \hline
       $C_\gamma$ for $B_0=10$~T & $0.50$ & $0.36$ & $0.25$ \\ \hline
    \end{tabular}
   \caption{CASH-II optimal unloaded quality factor $Q_0$, number of frequency shifts $N$, and setup sensitivity for $C_\gamma$ for various magnetic field strengths and total measurement times.}
    \label{tab:tab2}
\end{table}
In the experiment, the quality factor will be higher due to lower temperatures, but the loaded quality factor can be slightly adjusted by coupling between the antenna and the cavity. 
The number $N$ of frequency shifts as well as the setup sensitivity to $C_\gamma$ for two benchmark values of the magnetic field, 7~T and 10~T, are presented in Tab.~\ref{tab:tab2}.  The magnetic field necessary for the benchmark DSFZ coupling $C_\gamma=0.75$ for 1 year of observation is $3.3$ T. The time for a measurement at a fixed frequency is $t=1354$~s $\approx 22.6$~min, and $N_{\rm d.c.}=13.5$ dark counts for a single frequency measurement are expected per this run. The sensitivity for the upper limit of $95\%$ CL of $C_\gamma$ in the axion mass range of $m_a = (38 - 54)~\mu\mbox{eV}$ for $\tau=1$ year of operation is presented in Fig.~\ref{fig:sensitivity1}, labeled ``CASH-II'', and compared to other projects in Fig.~\ref{fig:sensitivity2}.

\section{Conclusions}
\label{sec:concl}
We propose a new CASH cavity haloscope experiment for the search of dark-matter axions with mass in the range $(38-54)\,\mu$eV, not covered yet by other projects. The main advantage of our proposed setup is the use of the Josephson junction based single-photon detectors, which significantly overcome the standard quantum limit. These high-precision detectors, as well as the ultimately low level of noise (not more than a single false switching event in 100~s) at ultra-low, $(10-20)$~mK, temperatures, provide for the sensitivity to halo axions reaching the DFSZ benchmark in one year of operation even with a $3.3$ T magnet. In a realistic setup with the field strength increased up to 10~T, the instrument would be sensitive to couplings twice smaller than the DFSZ value. Possible extensions of the CASH experiment may include multi-cell cavities, providing for a wider accessible axion mass range. Without the magnetic field, this haloscope is capable of searching for dark matter formed by new light particles of another kind, as dark photons.

\section*{Acknowledgments}
The research was partially supported by the National Center of Physics and Mathematics, directions No.\,5 and 8. The measurements of JJ SPD characteristics were supported by RSF
project 21-79-20227.

\bibliography{bibl}
\end{document}